\documentclass[10pt, conference, compsocconf]{IEEEtran}
\usepackage{epstopdf}

\usepackage{graphicx}
\usepackage{algorithm}
\usepackage[noend]{algpseudocode}
\usepackage{subfigure}
\usepackage{graphics}
\usepackage{tabularx, booktabs}
\newcolumntype{Y}{>{\centering\arraybackslash}X}
\newcommand*\Let[2]{\State #1 $\gets$ #2}
\algrenewcommand\algorithmicrequire{\textbf{Input:}}
\algrenewcommand\algorithmicensure{\textbf{Globals:}}
\usepackage{url}
\usepackage{hyperref}
\usepackage{breakurl}
\usepackage[font=normal,labelfont=bf]{caption}
\def\IEEEbibitemsep{0pt plus .6pt}

\begin{document}

\title{Exploiting Path Diversity in Datacenters Using MPTCP-aware SDN}

\author{\IEEEauthorblockN{Savvas Zannettou, Michael Sirivianos, Fragkiskos Papadopoulos}
\IEEEauthorblockA{Cyprus University of Technology\\
Limassol 3036,Cyprus\\
sa.zannettou@edu.cut.ac.cy, \{michael.sirivianos,f.papadopoulos\}@cut.ac.cy}}

\maketitle

\begin{abstract}
Recently, Multipath TCP (MPTCP) has been proposed as an alternative transport approach for datacenter networks. MPTCP provides 
the ability to split a flow into multiple paths thus providing better performance and resilience to failures.
Usually, MPTCP is combined with flow-based Equal-Cost Multi-Path Routing (ECMP), which uses random hashing to split the 
MPTCP subflows over different paths. However, random hashing can be suboptimal as distinct subflows may end up 
using the same paths, while other available paths remain unutilized.

In this paper, we explore an MPTCP-aware SDN controller that facilitates an alternative routing mechanism for the MPTCP subflows. 
The controller uses packet inspection to provide deterministic subflow assignment to paths. 
Using the controller, we show that MPTCP can deliver significantly improved performance when connections are not 
limited by the access links of hosts.  To lessen the effect of throughput limitation due to access links, we also 
investigate the usage of multiple interfaces at the hosts. We demonstrate, using our modification of the MPTCP 
Linux Kernel, that using multiple subflows per pair of IP addresses can yield improved performance in multi-interface settings.
\end{abstract}


\IEEEpeerreviewmaketitle

\section{Introduction}
Modern datacenters are responsible for executing data and computation intensive applications that often generate large flows. Providing high throughput to such applications and avoiding performance degradation due to bottlenecks created inside the network is of paramount importance. To this end, earlier studies have proposed different datacenter topologies such as FatTree~\cite{al2008scalable}, BCube~\cite{guo2009bcube}, VL2~\cite{greenberg2009vl2} and Jellyfish~\cite{singla2012jellyfish}, which aim to provide high aggregate throughput. Furthermore, Multipath TCP (MPTCP) ~\cite{ford2013tcp} was recently proposed as a new transport approach for datacenters~\cite{raiciu2011improving}. MPTCP outperforms regular TCP in terms of performance and robustness. Intuitively, this is because MPTCP is able to strip data to multiple paths inside the network, by creating multiple subflows, while offering load balancing by sending more data to the least congested paths~\cite{mptcp_kernel_implementation}. 

A key aspect that affects MPTCP performance is the routing mechanism of the subflows. Currently, the most 
prominent and widely deployed routing mechanism in datacenters is a flow-based variant of Equal-Cost Multi-Path 
Routing (ECMP)~\cite{hopps2000analysis}. Flow-based ECMP uses random hashing to uniformly split the subflows over different shortest paths. However, random hashing works suboptimally as subflows may end up using the same paths, while available paths remain unutilized. For example, with two subflows and two available paths, there is a $50\%$ probability to assign the two subflows to the same path. Another limitation is that the available path diversity may not be well-exploited as only shortest paths are considered.

In this paper, we explore an MPTCP-aware Software-Defined Networking (SDN) controller for routing the MPTCP subflows. The use of an SDN controller provides an ideal environment for implementing more efficient routing mechanisms for the subflows, as the controller maintains a global view of the network~\cite{al2010hedera}. The controller can calculate various sets of paths between two hosts, such as shortest paths, $k$-shortest paths, $k$-edge-disjoint paths, etc., thus better exploiting the available path diversity in a given datacenter topology. 

It is tempting to have more subflows per MPTCP connection to better exploit path diversity. However,
the creation and maintenance of a large number of subflows imposes extra overheads to the end hosts, which 
require larger buffers to cope with reordering, and higher CPU utilization due to the increased usage of the MPTCP scheduler. More importantly however, in an SDN environment a larger number of subflows imposes extra overheads to the network, by requiring a larger number of rules to be installed at the switches and by increasing the load at the SDN controller (see discussion in Section~\ref{sec:discussion}). It is thus desirable to have as few subflows as possible without sacrificing performance. Given a set of paths, the proposed SDN controller can better utilize these paths by deterministically assigning MPTCP subflows to them. As we show in this paper, this can result in near-optimal performance without the need for as many subflows as in random-based approaches.  

To facilitate deterministic assignment of subflows to paths, our controller performs packet inspection to extract the MPTCP options and stores information regarding MPTCP connections. We compare this MPTCP-aware approach to the random-based approach used in ECMP. We find that when MPTCP connections are not limited by the host access links, the MPTCP-aware approach provides significant performance gains, and is able to achieve near-optimal performance with fewer subflows. When MPTCP connections are limited by the host access links, the MPTCP-aware and the random-based approaches perform similarly. This is because the bottlenecks  lie at the endpoints of the connections, and the better exploitation of paths inside the network yields no benefits. 
To improve performance in such cases, we explore the use of multiple interfaces at the hosts in conjunction with our MPTCP-aware approach. Using a dual-homed variant of the Jellyfish topology~\cite{singla2012jellyfish} and our MPTCP Linux Kernel modification~\cite{mptcp_patch}, we demonstrate that multiple subflows per pair of IP addresses can significantly improve performance.  Our modification has been included in the latest version (v0.90) of the MPTCP Linux Kernel~\cite{mptcp_kernel_implementation}. Further, we make our SDN controller publicly available~\cite{mptcp_aware_controller_link}.

The rest of the paper is organized as follows. In Section \ref{related} we review the most relevant earlier work and in Section \ref{background} we provide the necessary MPTCP background. In Section \ref{overview_section} we present the design of our system, and in Section \ref{evaluation} we perform our evaluation. A discussion follows in Section~\ref{sec:discussion}, and we conclude in Section \ref{conclusion_section}.

\section{Related Work} 
\label{related}

Raiciu et al.~\cite{raiciu2011improving} demonstrated that the use of MPTCP in datacenters can be beneficial in terms of performance and robustness. The authors used flow-based ECMP to route the MPTCP subflows, and proposed using a dual-homed variant of the FatTree topology to improve performance. In this paper, we consider an MPTCP-aware SDN approach for routing the subflows, which can exploit more efficiently the available paths. Furthermore, using our modification of the MPTCP Linux Kernel at the hosts, we explore a dual-homed variant of the Jellyfish topology that provides more path diversity than FatTree. 

There have been a few recent studies demonstrating the merits of integrating MPTCP with SDN. The most relevant to our work is the one by Sandri et al.~\cite{sandri2015benefits}, who proposed an SDN controller that distributes subflows belonging to the same MPTCP connection over distinct disjoint paths. The controller requires storing the network topology for each MPTCP connection, and cannot exploit settings where hosts have multiple interfaces. Specifically, the controller treats each subflow between two hosts the same, irrespectively of the subflow's source-destination IP addresses (interfaces). Our controller, does not store a topology for each MPTCP connection, but only a set of paths, which is a significantly smaller subset of the topology. More importantly, our controller is designed for multi-interface settings, where we show, using our MPTCP Linux Kernel modification, that the creation of multiple subflows per pair of IP addresses can significantly improve performance. Further, Sandri et al.~\cite{sandri2015benefits} considered only small toy topologies where host access links were not bottlenecks. Here, we focus on datacenter topologies that have specific structural characteristics, and explore the usage of multiple interfaces in cases where access links can be bottlenecks. 

Detal et al.~\cite{detal2013revisiting} demonstrated the performance improvements of a deterministic routing approach in datacenters. Specifically, the authors proposed a routing approach where end hosts are able to select packet header values in order to force the selection of a specific path. To achieve this, the authors implemented a user space extension that enables end hosts to select TCP header values that determine the desired path. Furthermore, to facilitate the deterministic approach the authors implemented a new path selection mechanism on switches (instead of the hash function in typical flow-based ECMP). Taken altogether, the approach in~\cite{detal2013revisiting} requires additional modifications on end hosts and switches. On the contrary, our approach can be deployed on typical OpenFlow-enabled networks by using our SDN controller~\cite{mptcp_aware_controller_link} without the need of extra modifications on end hosts or switches. 

Agache et al.~\cite{agache2015increasing} study the use of multiple host interfaces in datacenters. The authors do not use multihoming (hosts being connected to more than one switch) because it increases the overall costs due to extra switches. Instead, the authors proposed GRIN, a datacenter architecture based on VL2~\cite{greenberg2009vl2}, which interconnects free interfaces between hosts. Intuitively, a host in GRIN can utilize another host's interface when that interface is not being utilized, thus opportunistically increasing its bandwidth. By design, GRIN provides significant performance improvements only when a considerable percentage of hosts are not heavily utilizing their interfaces. On the other hand, we demonstrate that our dual-homed solution (DH-Jellyfish) can achieve significant performance improvements when all hosts are fully utilizing their interfaces. Note that our approach can deliver the full merits of multihoming without increasing costs (same number of switches).

\section{MPTCP Background} 
\label{background}

In this section we present the required MPTCP background for the design of our SDN controller. 
More information about MPTCP can be found in~\cite{ford2013tcp}. 

MPTCP presents a regular TCP interface to applications, while in fact it spreads data across several subflows. 
Each MPTCP connection has a unique identifier, called $token$, which is used for the association/authentication of new subflows. Fundamentally, a subflow is established similarly to a TCP connection with the difference that the handshake contains MPTCP-specific options such as MP\_CAPABLE and MP\_JOIN. The MP\_CAPABLE option is used only during the initial subflow to verify whether the remote host is MPTCP-enabled. The MP\_JOIN option is used for the establishment of each additional subflow and its association with the MPTCP connection.
\begin{figure}[ht]
\centering
\includegraphics [width=2.8in, height=1.7in]{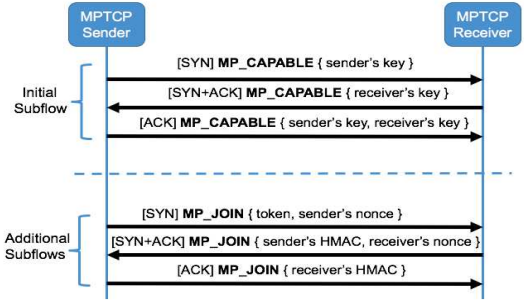}
\caption{Subflow establishment in MPTCP.}
\label{connection_establishment}
\end{figure}

Figure~\ref{connection_establishment} depicts the subflow establishment of an MPTCP connection. For the initial subflow, the hosts perform a handshake with the MP\_CAPABLE option that contains randomly generated keys. 
These keys are used for the calculation of the token (cryptographic hash of the keys). For the establishment of additional subflows, the hosts perform a handshake with the MP\_JOIN option as follows. The sender sends a SYN packet with the token and a random nonce. (The MP\_CAPABLE and MP\_JOIN SYN packets are the subflow setup packets). Upon reception of the packet, the receiver responds with a SYN+ACK packet containing its own nonce and a calculated HMAC code of the sender's nonce. Finally, the sender sends an ACK packet that contains the calculated HMAC of the receiver's nonce. The HMACs are calculated using the keys exchanged during the MP\_CAPABLE handshake, and the nonces exchanged in the MP\_JOIN handshakes.

\section{System Overview} 
\label{overview_section}

\begin{figure}[ht!]
\centering
\includegraphics[width=\linewidth,height=1.6in]{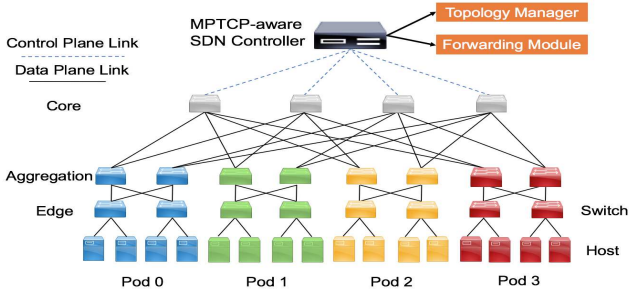}
\caption{System Overview}
\label{system_overview}
\end{figure}
Figure~\ref{system_overview} provides a general overview of the system we consider. The transport protocol is MPTCP with the extra modification described in Section~\ref{mptcp_path_manager_section}. 
The system takes as input an underlying datacenter topology, where each switch is connected to the SDN controller through control plane links---for clarity, only four such links are shown. The controller is responsible for the calculation of paths (consisting of data plane links), and the installation of the appropriate OpenFlow~\cite{mckeown2008openflow} rules to the switches. When a packet arrives at a switch, the rules tell the switch where to forward the packet. When a packet of a subflow arrives at a switch and a rule does not exist, the packet is redirected to the controller. The controller calculates the path for the subflow and installs rules to all the switches that are in the calculated path. Subsequent packets of the same subflow traverse the same path and avoid redirection to 
the controller. Below, we provide a detailed description of the topology and the system's components.

\subsection{Topology}

The topology plays a major role in the performance of a datacenter. In this paper, we consider the 
FatTree~\cite{al2008scalable} and Jellyfish~\cite{singla2012jellyfish} topologies. Both topologies are built using 
commodity switches with uniform link capacities, but they have some fundamental differences. 

FatTree is probably the most popular \emph{structured} datacenter topology, aiming to deliver high aggregate 
throughput while keeping costs low. Considering a $k$-ary FatTree topology with three levels of switches, each 
one consisting of $k$ ports, the topology interconnects $k^3/4$ hosts. There are $k$ pods, each containing two layers of $k/2$ switches (edge and aggregation levels). Each switch at the edge level connects $k/2$ hosts. For hosts that reside in different pods there are $k^2/4$ shortest paths and $k/2$ edge-disjoint paths. Figure \ref{system_overview} shows a FatTree topology with $k=4$.

Jellyfish is the most prominent \emph{randomly constructed} topology. It can hold more hosts than FatTree while 
maintaining the same performance~\cite{singla2012jellyfish}. It also provides the flexibility to define the number of 
switches, ports per switch and hosts at will. Each host has one interface that is connected to one switch. Hosts 
are assumed uniformly distributed among the switches, i.e., the number of hosts per switch is 
$\lceil \frac{N}{S} \rceil$, where $N$ is the total number of hosts and $S$ the total number of switches. As a result, all switches have a similar number of free switch ports devoted for the core of the network. To construct the core, pairs of switches are connected at random until all ports in all switches are used. Since Jellyfish is a random graph, there is no prior knowledge of the available shortest and edge-disjoint paths between hosts as in FatTree. 

\emph{Dual-homed Jellyfish.} We also propose and explore a dual-homed variant of Jellyfish (DH-Jellyfish). DH-Jellyfish uses the same equipment as Jellyfish but each host has two interfaces. The construction of DH-Jellyfish is the same with Jellyfish with the following addition. For the second interface at each host, we randomly select a switch and connect the host, given that: (i) the host is not already connected to the switch; and (ii) the switch has not been already connected $\lceil \frac{N}{S} \rceil$ times. 
The second constraint ensures that we uniformly connect the extra interfaces to the switches. DH-Jellyfish sacrifices some path diversity in the core (as switches have fewer available ports to interconnect) in favor of increased bandwidth at the endpoints. As we show, DH-Jellyfish can offer significant performance improvements compared to Jellyfish in our MPTCP-aware approach. 

\subsection{Controller} 
\label{controller_subsection}

We have based our controller implementation on Floodlight~\cite{floodlight}. It consists of two main components: the Topology Manager (TM) and the Forwarding Module (FM). The TM calculates and provides sets of paths between host interfaces to the FM. The FM is responsible for the selection of paths and the installation of the appropriate OpenFlow rules to the switches. These rules consist of header values to match packets and an associated action to apply to matching packets. Specifically, the matching header values in our case are source-destination IP addresses and port numbers, which together identify packets belonging to the same subflow. The action is the port of the switch to forward the matching packets. Below, we describe the TM and FM components in detail.

\subsubsection{Topology Manager (TM)} 
\label{tm}

The TM has an up-to-date global view of the topology. This is obtained by messages exchanged between the SDN 
controller and the switches (see the LinkDiscoveryManager module in~\cite{floodlight}). The TM is queried by the FM whenever a set of paths between source-destination interfaces (IP addresses) is needed. The TM uses the Depth First Search (DFS) graph traversal algorithm and finds all available paths, whose lengths do not exceed a certain hop-count threshold. Subsequently, the TM filters the obtained set of paths to extract one of the following subsets, 
which is returned to the FM: (i) \emph{shortest paths}, which contains all the paths that have the same hop-count as the shortest path; (ii) \emph{$k$-shortest paths}, which contains the first $k$ paths in increasing hop-count order; and (iii) \emph{$k$-edge-disjoint paths}, which contains the first $k$ paths that do not share any edges in 
increasing hop-count order.

The selection of a good set of paths is crucial for the performance of MPTCP. The set of paths should exploit 
the available path redundancy in the topology, while being as small as possible. If these two conditions are met, then with a deterministic subflow assignment to paths we should be able to maximize MPTCP's performance with a small number of subflows. As mentioned, besides decreasing overheads at the end hosts, a small number of subflows is also desirable as it will minimize the number of rules installed at the switches and the load (redirections of subflow setup packets) to the FM.

\subsubsection{Forwarding Module (FM)}
\label{fm}

The FM can perform either deterministic or random assignment of subflows to paths. In the latter, the FM 
randomly selects a path from the obtained set of paths for the source-destination IP addresses of the subflow, 
and assigns the subflow to the path. If the set of paths is the set of shortest paths, then this approach is equivalent to flow-based ECMP.  

In the deterministic approach, each subflow is assigned to a different path, which belongs to the obtained set of paths for the source-destination IP addresses of the subflow. This approach is {\it MPTCP-aware} because it requires the FM  to extract the MPTCP options from the subflow setup packets to match a subflow to an existing MPTCP connection (if there is one). The goal is to avoid assigning more than one subflows of the same connection to a single path. If the number of subflows is larger than the number of available paths, the subflows are deterministically assigned to paths in a uniform manner.  

The FM maintains two hashtables: $pathCache$ and $flows$. $pathCache$ is used to cache sets of paths 
between source-destination IP addresses, which are obtained by querying the TM. Specifically, when a new subflow setup packet arrives, the FM first queries $pathCache$ to see if a set of paths corresponding to the source-destination IP addresses of the subflow exists. If it exists, the FM loads this set of paths to the $flows$ hashtable and maps it to the requested MPTCP connection. Otherwise, the FM queries the TM and stores the obtained set of paths to both $pathCache$ and $flows$. The entries in $pathCache$ are set to expire every $60$ minutes, so that paths can get refreshed by querying the TM. This is a reasonable time interval in which datacenter topologies can be considered almost static. Indeed, Gill et al.~\cite{gill2011understanding} showed that switches in datacenters have median time between failures in the order of multiple hours. In case of failures within the $60$ minutes interval, failed paths will not be utilized but, MPTCP will be able to achieve acceptable performance as it will only send traffic through the good paths.

The $flows$ hashtable contains an entry for each MPTCP connection. Each such entry consists of subentries 
for each pair of source-destination IP addresses (interfaces) used by the subflows of the connection. Each such subentry, called $IPentry$, caches the obtained set of paths for a source-destination IP address pair, as well as the current assignment of subflows to paths in the set. Using this information, the FM is able to keep track of the available paths it can choose from for a new subflow. The entries for an MPTCP connection are expired from $flows$ after $5$ seconds, as the cached information is useful only until the establishment of all the connection's subflows, which happens at the beginning of the connection~\cite{mptcp_kernel_implementation}.
\begin{algorithm}[t]
\small
  \caption{Forwarding Module pseudocode
    \label{alg:forwarding}}
  \begin{algorithmic}[1]
    \Require{$p$ is the subflow setup packet received at the controller}
      \Let{$IPs$}{Extract source/destination IP addresses from $p$}  
      \Let{$ports$}{Extract source/destination port numbers from $p$}       
           \Let{$type$}{Extract MPTCP option from $p$}
      \If{$type == MP\textunderscore{}CAPABLE$}
      	    \State Store $IPs$ in $primaryIPs$
          \State Query $pathCache$/TM to find the shortest path for $IPs$
          \State If TM queried, update $pathCache$
          \Let{$path$}{shortest path}
      \EndIf
      \If{$type == MP\textunderscore{}JOIN$}
          \Let{$token$}{Extract MPTCP token from $p$}
          \If{$token$ does not exist in $flows$}
              \State Create a new $entry$ in $flows$ using the $token$
              \State Query $pathCache$/TM to get set of paths for $IPs$
              \State If TM queried, update $pathCache$
              \State Create a new $IPentry$ in $entry$ using $IPs$
              \Let{$path$}{Get next path from $IPentry$}
              \State Update $IPentry$
          \Else
             \Let{$entry$}{$flows[token]$}
             \If{$IPs$ exists in $entry$}
           		\Let{$IPentry$}{$entry[IPs]$}
          		\Let{$path$}{Get next path from $IPentry$}
          		\State Update $IPentry$
              \Else
      		       \State Query $pathCache$/TM to get set of paths for $IPs$
          	       \State If TM queried, update $pathCache$     
               	\State Create a new $IPentry$ in $entry$ using $IPs$
			\Let{$path$}{Get next path from $IPentry$}
                   \State Update $IPentry$
              \EndIf
           \EndIf
         \EndIf
         \State Install rules for the $path$ to the switches using $IPs$ and $ports$
    \end{algorithmic}
\normalfont
\end{algorithm}

Algorithm~\ref{alg:forwarding} shows the detailed pseudocode of the MPTCP-aware FM. When the connection setup 
packet of a new subflow arrives at the first switch of the network, the switch forwards the packet to the FM as there are no forwarding rules installed yet at the switches for the packets of the subflow. Upon reception of the packet, the FM extracts the source-destination IP addresses ($IPs$) and port numbers ($ports$) (lines~1,2). It also extracts the MPTCP options (line~3), and distinguishes the cases MP\_CAPABLE and MP\_JOIN. If the option is MP\_CAPABLE (initial subflow), the FM stores the $IPs$ in a $primaryIPs$ hashtable (lines~4,5). It then finds the shortest path between these $IPs$ by querying $pathCache$ or the TM, and assigns this path to the subflow (lines~6-8). The FM always assigns to the initial subflow the shortest path and keeps track of its $IPs$ in $primaryIPs$---the reason for this is to avoid assigning the same path to additional subflows of the connection with the same $IPs$.

If the option is MP\_JOIN (additional subflow), the FM extracts the token from the packet, which identifies the 
existing MPTCP connection (lines~9,10). If the token does not exist in $flows$, which occurs only if this is the 
first additional subflow, the FM creates a new $entry$ in $flows$ using the token (lines~11,12). Subsequently, the FM 
queries $pathCache$ or the TM to get a set of paths for the $IPs$, and creates a new $IPentry$ in $entry$ for these $IPs$ (lines~13-15). This $IPentry$ stores the obtained set of paths. One of these paths is then assigned to the subflow, and the $IPentry$ is updated to indicate this assignment (lines~16,17). If the token exists in $flows$, the corresponding $entry$ is retrieved (line~19), and the FM checks if the $IPs$ exist in $entry$ (line~20). If they exist, the FM retrieves the corresponding $IPentry$, from where the next available path is assigned to the subflow (lines~21,22). The FM then updates the $IPentry$ as before (line~23). If the $IPs$ do not exist in $entry$, the FM queries $pathCache$ or the TM to get a set of paths for the $IPs$, and creates a corresponding $IPentry$ in $entry$ as before (lines~25-27). It then assigns a path to the subflow and updates $IPentry$ (lines~28,29). After the selection of a path, the FM installs forwarding rules for the subflow, which is identified by its $IPs$ and $ports$, on all the switches that belong to the assigned path (line~30). 

We note that the switches delete a rule from their table if they do not receive a packet matching that rule for a specified time interval. The default value of this interval in Floodlight is 5 seconds~\cite{floodlight}. We also note that the usage of $pathCache$ and $flows$ significantly reduces the processing time of the subflow setup packets at the controller. By processing time we mean the time interval between the reception of a subflow setup packet at the controller and the installation of the corresponding rules at the switches. This time includes all the required processing from the FM and the TM if required. In our evaluation, the average processing time for subflow setup packets that do not require querying the TM is around $1$ms vs. $280$ms for setup packets that require querying the TM. We note that the path calculation in our TM implementation is currently not performance-optimal. Specifically, our TM calculates all possible paths between two hosts within a given hop-count threshold, and many of the calculated paths may never be used. By optimizing the path calculation process, the processing time for subflow setup packets that require querying the TM can decrease. Such optimization however is beyond the scope of the present work.

\begin{table*}
\begin{tabular}{cc}
    \begin{minipage}[t]{.35\linewidth}
    
\begin{tabularx}{\textwidth}{c *{6}{Y}}
\toprule

 & \multicolumn{2}{c}{M-Disjoint(4)}  
 & \multicolumn{2}{c}{R-Shortest(16)}\\

 Subflows & PT & UT  & PT & UT \\
\midrule
1        & 48.5\%  		 & 45.1\%         &49.1\%         & 44.1\%                           \\ \hline
2        & 69.7\%         & 58.7\%		 & 67.1\%         & 55.8\%                  \\ \hline
3        & 82.7\%         & 63.9\%       & 77.7\%         & 60.5\%                  \\ \hline
\textbf{4}        & \textbf{90.0\%}          & \textbf{64.3\%} & \textbf{83.8\%}         & \textbf{62.4\%}                  \\ \hline
5        & 90.5\%          & 65.0\%	& 87.8\%         & 63.2\%                  \\ \hline
6         & 90.8\%          & 65.4\% & 89.9\%         & 64.1\%                 \\ 

\bottomrule
\end{tabularx}

\caption{Average MPTCP throughput in FatTree (\% of optimal).}
\label{ft_table}
    \end{minipage} &

    \begin{minipage}[t]{.65\linewidth}
\captionsetup{width=0.99\textwidth}

\begin{tabularx}{\textwidth}{c*{9}{Y}}
\toprule
         & \multicolumn{2}{c}{M-Disjoint(8)}
         & \multicolumn{2}{c}{M-Shortest(8)}  
         & \multicolumn{2}{c}{R-Disjoint(8)} 
         & \multicolumn{2}{c}{R-Shortest(8)} \\
         
Subflows & PT                  & UT                 & PT                    & UT                    & PT                  & UT                 & PT                    & UT                    \\
\midrule

1         & 67.7\%                & 45.2\%	 	& 66.7\%                & 48.2\%               		& 49.5\%              & 44.3\%                            & 54.2\%              & 45.3\%             \\ \hline
2        & 83.6\%                & 58.5\% 		& 81.9\%                & 60.5\%               		& 66.1\%              & 58.8\%                            & 69.7\%              & 57.3\%              \\ \hline
3         & 93.1\%                & 59.5\%		& 89.8\%                & 62.4\%                 		& 77.1\%              & 58.9\%                          & 79.5\%              & 61.2\%              \\ \hline
\textbf{4}     & \textbf{95.6\%}                & \textbf{66.1\%}  & \textbf{93.6\%}                  & \textbf{63.0\%}    & \textbf{81.9\%}              & \textbf{63.6\%}                            & \textbf{84.8\%}              & \textbf{62.0\%}                            \\ \hline
5         & 96.0\%                & 66.7\%		& 95.1\%                & 63.2\%               		& 86.2\%              & 63.6\%                            & 87.4\%              & 62.6\%              \\ \hline
6         & 96.0\%                & 67.2\%        & 95.4\%                & 63.4\%					& 88.4\%              & 63.5\%                            & 89.9\%              & 63.0\%                             \\ 
\bottomrule
\end{tabularx}$  $
\caption{Average MPTCP throughput in Jellyfish (\% of optimal).}
\label{jelly_table}
    \end{minipage} 
\end{tabular}
\end{table*}

\begin{figure*}
\centerline{
\subfigure[FatTree.]{\includegraphics [width=2in, height=1.4in]{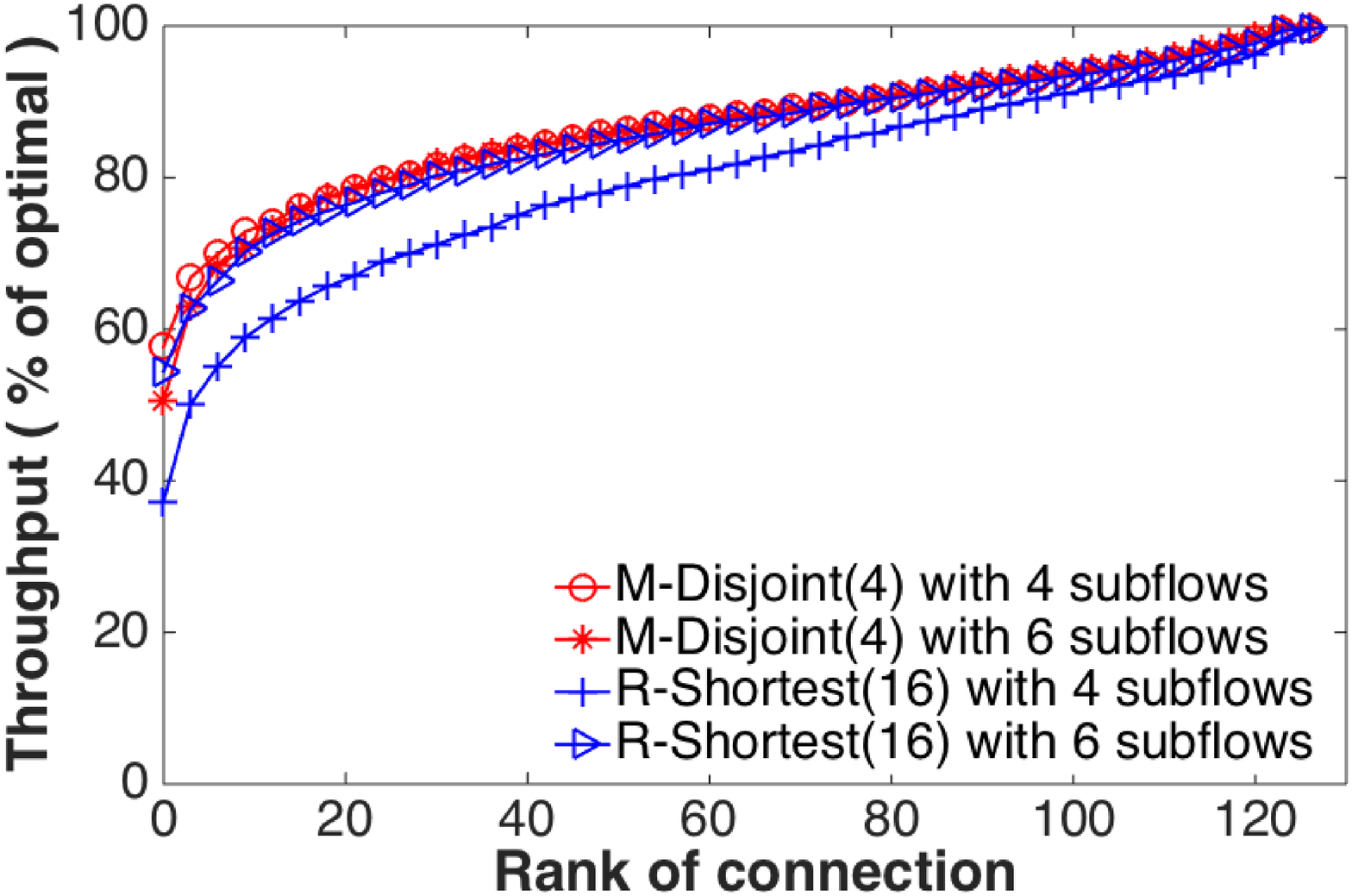}}
\subfigure[Jellyfish - Disjoint paths.]{\includegraphics [width=2in, height=1.4in]{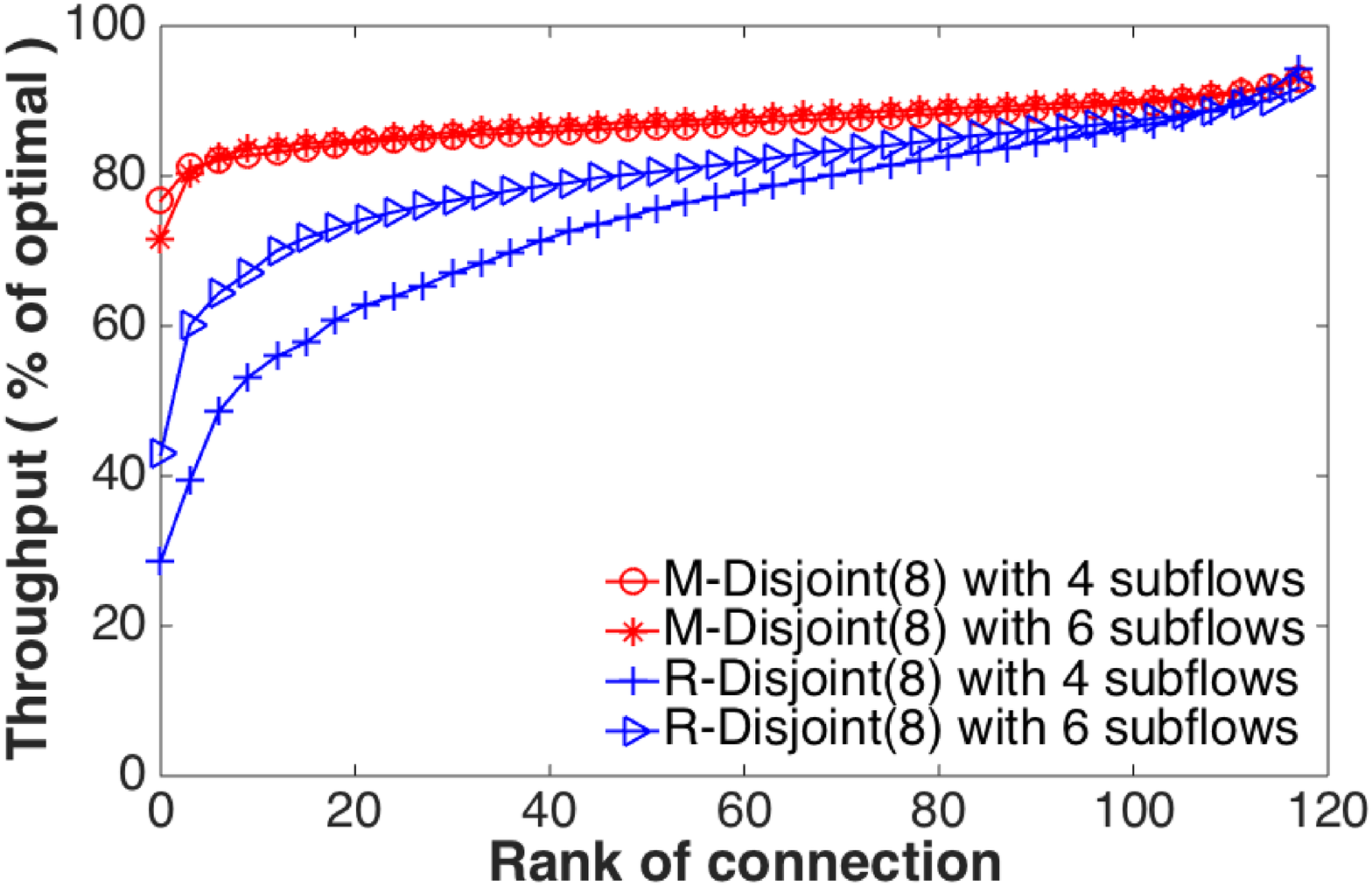}}
\subfigure[Jellyfish - Shortest paths.]{\includegraphics [width=2in, height=1.4in]{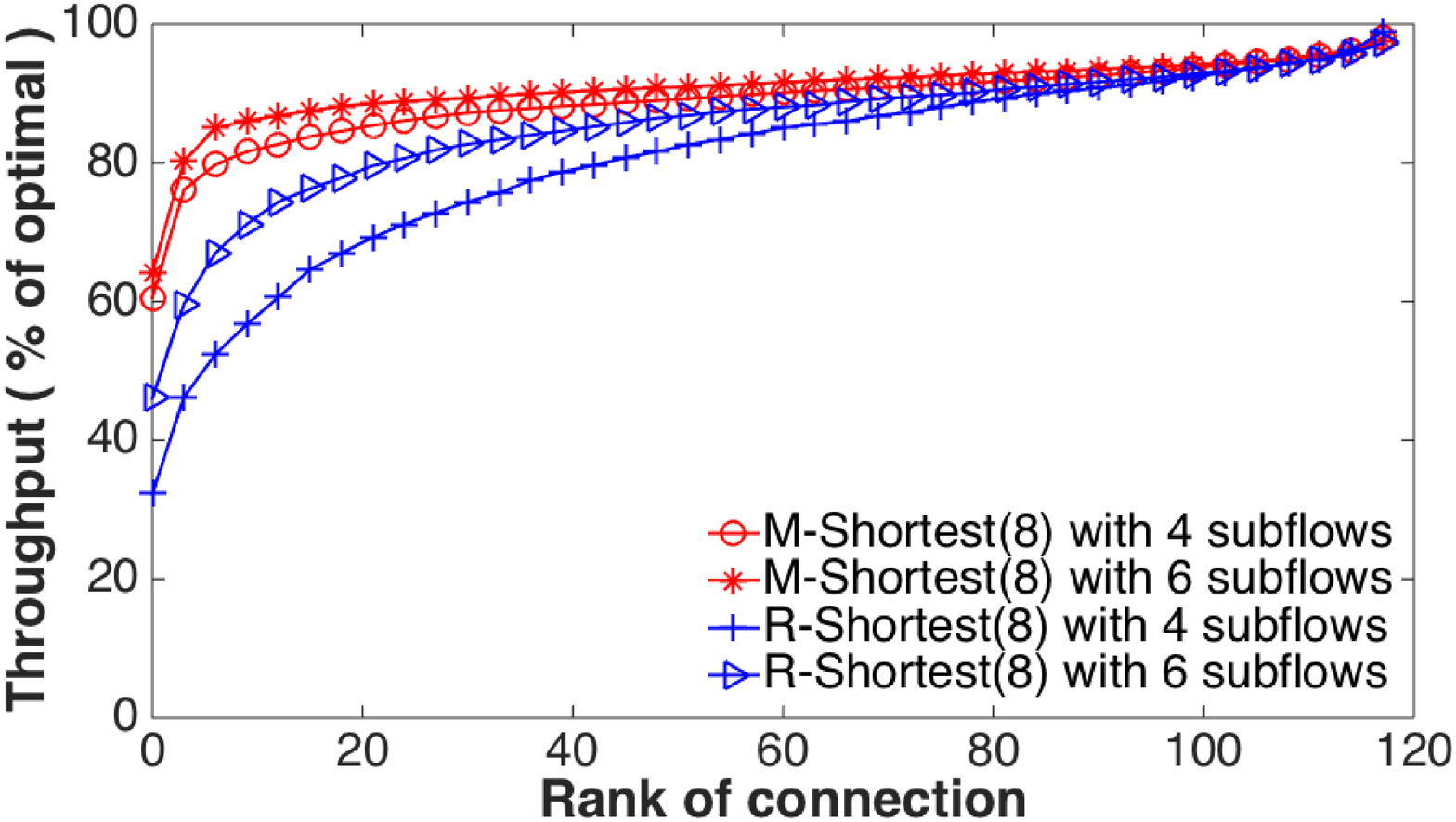}}}
\caption{Distribution of MPTCP throughput.}
\label{ft_jelly_rank}
\end{figure*}
\subsection{An improved MPTCP path manager} 
\label{mptcp_path_manager_section}

The component responsible for the creation of subflows in the MPTCP Linux Kernel is called path manager. 
The implementation offers two path managers: \emph{fullmesh} and \emph{ndiffports}~\cite{mptcp_kernel_implementation}. 
However, prior to our contribution these path managers did not offer the functionality of creating more than one 
subflow per pair of source-destination interfaces (IP addresses), when MPTCP uses multiple interfaces at 
the source and/or destination. This limitation was preventing MPTCP from exploiting multiple paths between the 
same pair of IP addresses in multi-interface settings.

Motivated by the above observation, we have extended the \emph{fullmesh} path manager, which now offers the flexibility to 
arbitrarily change the number of subflows per pair of IP addresses in multi-interface settings. 
Our patch~\cite{mptcp_patch} has been included in the latest release (v0.90) of the MPTCP Linux Kernel~\cite{mptcp_kernel_implementation}. 
In Section \ref{evaluation_section}, we demonstrate the merits of the extended~\emph{fullmesh} in the DH-Jellyfish topology.

\section{Evaluation} 
\label{evaluation}

\subsection{Setup}
\label{experimental_section}

For our evaluation we used the Mininet emulator~\cite{mininet_paper} and the MPTCP Linux Kernel 
Implementation v.0.90. For the emulation of the switches we used Open vSwitch~\cite{pfaff2009extending}. 
The topologies, traffic patterns, and subflow routing mechanisms used are described below.

\textbf{Topologies.} We emulated an 8-ary FatTree topology having 128 hosts and 80 switches (see Fig.~\ref{system_overview} for a 4-ary FatTree). By construction, in this topology there are 16 shortest paths and 4 edge-disjoint paths between hosts that reside in different pods. We also emulated Jellyfish and DH-Jellyfish topologies, consisting of 120 hosts and 60 switches each. In all cases, all links have the same capacity.

\textbf{Traffic patterns.} We considered two different types of traffic patterns: (i) the unconstraint traffic matrix (UT); and (ii) the permutation traffic matrix (PT). In both cases, we randomly select $N$ source-destination pairs of hosts, where $N$ is the number of hosts in the topology. Then, we establish long-lived MPTCP connections between all selected pairs of hosts---all connections are initiated at the same time. In UT, a host can participate in any number of connections, meaning that the host access links can be bottlenecks. In PT, we have the additional selection constraint that no host can participate in more than one connection. This ensures that the host access links are not bottlenecks for the connections, which can share links only inside the core of the topology. 

\textbf{Subflow routing.} For subflow routing we consider both the MPTCP-aware approach (depicted as M) as well as the random-based approach (depicted as R), which are implemented in our FM. We also consider the cases where the source-destination paths are shortest paths, $k$-shortest paths, and $k$-edge-disjoint paths.

\subsection{Results}
\label{evaluation_section}

We use the following evaluation metrics: (i) the average throughput over all MPTCP connections; and (ii) 
the distribution of the MPTCP throughput, represented by the ranking of the individual connection throughputs, from 
the worst throughput to the best throughput. The latter metric provides details about the performance of the slowest connection. This metric is important in datacenters where jobs create multiple collaborating workers because the overall performance is dictated by the performance of the slowest worker. All metrics are in the form of percentages compared to the optimal throughput, which is the capacity of the host interface in FatTree and Jellyfish, and the sum of capacities of the host interfaces in DH-Jellyfish. For example, if we assume uniform link capacities of 1Gb/s in the network topologies, then the optimal throughput is 1Gb/s for the FatTree/Jellyfish topology and 2Gb/s for the DH-Jellyfish topology. We say that the achieved throughput is \emph{near-optimal} if it is not smaller than 90\% of the optimal. All results are averaged over 10 different runs.


\textbf{FatTree.} Table~\ref{ft_table} shows the average MPTCP throughput in the FatTree topology with the MPTCP-aware (M) and random (R) approaches. The random approach uses all available (16) shortest paths between hosts, akin to flow-based ECMP. The MPTCP-aware approach uses all (4) edge-disjoint paths. For the PT, the MPTCP-aware approach requires only 4 subflows to achieve near-optimal performance, while the random approach requires 6 subflows. That is, the random approach requires 50\% more subflows and 300\% more paths to achieve near-optimal performance. Figure~\ref{ft_jelly_rank}(a)  shows the distribution of the MPTCP throughput with the PT. The worst throughput in the MPTCP-aware approach with 4 subflows is 57\% of the optimal, while in the random approach with the same number of subflows is 37\%. This corresponds to 54\% improvement in the MPTCP-aware approach. By further increasing the number of subflows, the random approach better utilizes the available shortest paths and its performance becomes similar to that of the MPTCP-aware approach. For the UT, several MPTCP connections have their throughput limited by the host access links. Because of this, the better exploitation of distinct paths inside the core of the topology yields no significant benefits. In this case, the MPTCP-aware and the random-based approaches perform similarly.

\textbf{Jellyfish.} Table~\ref{jelly_table} shows the average MPTCP throughput in Jellyfish. We consider the case 
where both MPTCP-aware and random-based approaches use 8-edge-disjoint paths, as well as the case where both approaches use 8-shortest paths. Figure~\ref{ft_jelly_rank}(b) shows the corresponding throughput distributions when the traffic pattern is the PT and 8-edge-disjoint paths are used---a similar behaviour holds for 8-shortest paths (Fig.~\ref{ft_jelly_rank}(c)). Our observations with the PT are the following. First, with only 4 subflows, the MPTCP-aware approach achieves near-optimal performance in both cases, while the random approach always performs worse (Table~\ref{jelly_table}). 
Second, with 4 subflows and 8-edge-disjoint paths, the worst throughput in the MPTCP-aware approach is 77\% of the optimal, whereas in the random-based approach is only 28\% (Fig.~\ref{ft_jelly_rank}(b)). This corresponds to 175\% improvement in the MPTCP-aware approach. Third, we note that significant performance benefits can also be achieved if k-shortest paths are used. Specifically, with only 4 subflows and 8-shortest paths, the worst throughput in the MPTCP-aware approach is 60\% of the optimal, while in the random-based approach is only 32\% (Fig. ~\ref{ft_jelly_rank}(c)). This corresponds to 88\% improvement in the MPTCP-aware approach. Finally, we observe that the performance gains are more pronounced than in FatTree (cf. Fig.~\ref{ft_jelly_rank}(a) vs. Fig.~\ref{ft_jelly_rank}(b)). This is because Jellyfish provides more path diversity, which we exploit (e.g., by using 8-edge-disjoint paths). However, similar to FatTree, when the traffic pattern is the UT, the MPTCP-aware and the random-based approaches perform similarly (Table~\ref{jelly_table}).

\textbf{DH-Jellyfish.} As can be seen in Tables~\ref{ft_table},\ref{jelly_table}, the average throughput with the UT 
is smaller than with the PT, in all cases. This is expected since connections in the UT can share access links that become bottlenecks. As explained, to mitigate this effect, hosts in DH-Jellyfish have one additional interface that can be used by MPTCP to send additional subflows. Note that in this dual-interface setting, if a source has the interfaces $a,b$ and the destination the interfaces $c,d$, the MPTCP connection will send subflows over all pairs of interfaces: $\{a,c\}, \{a,d\}, \{b,c\}, \{b,d\}$. That is, each connection uses 4 pairs of source-destination IP addresses. 

\begin{figure}
\centerline{
\subfigure[]{\includegraphics [width=1.77in, height=1.3in]{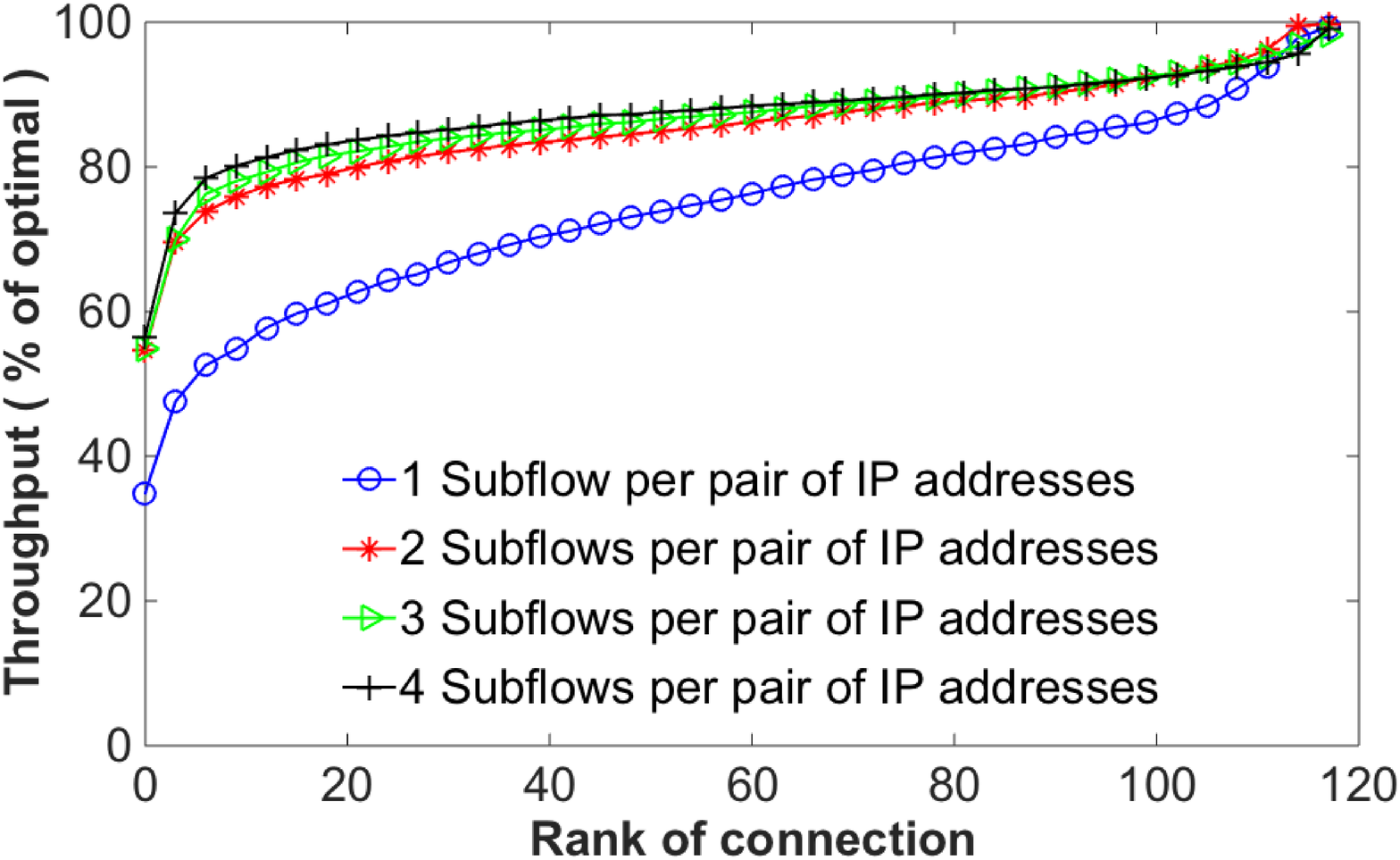}}\hfill
\subfigure[]{\includegraphics [width=1.77in, height=1.3in]{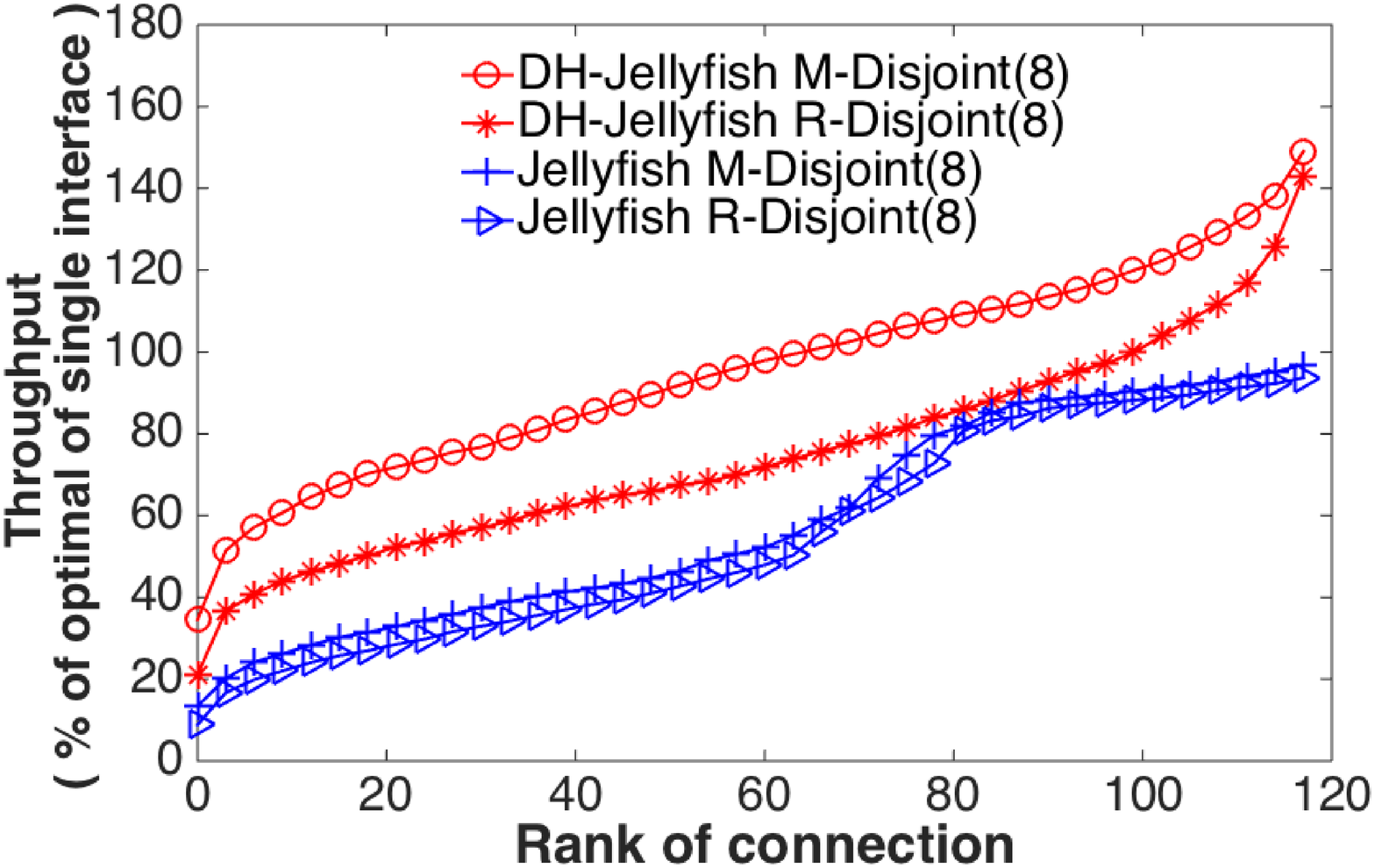}}}
\caption{(a) MPTCP throughput~vs.~number of subflows per pair of IP addresses in DH-Jellyfish; and 
(b) MPTCP-aware~vs.~random-based approach in DH-Jellyfish and Jellyfish.}
\label{dh_results}
\end{figure}

Figure~\ref{dh_results}(a) shows the distribution of the MPTCP throughput with 1, 2, 3 and 4 subflows per pair of IP addresses. The results correspond to the MPTCP-aware approach, which uses 8-edge-disjoint paths between source-destination IP addresses. To be able to send more than 1 subflow per pair of addresses, we used our modification of the MPTCP Linux Kernel described in Section~\ref{mptcp_path_manager_section}.  We observe that with 2 subflows per pair of addresses ($2\times4=8$ subflows in total) the throughput can be improved by up to 50\% compared to the case where 1 subflow per pair of addresses is used ($1\times4=4$ subflows in total). 
This is because we exploit the path redundancy that exists between each pair of IP addresses. We observe no additional gains by further increasing the number of subflows.

Figure~\ref{dh_results}(b) juxtaposes the performance of the MPTCP-aware and random-based approaches in DH-Jellyfish and Jellyfish. The y-axis in the plot is the throughput of each MPTCP connection normalized by the capacity of a single interface, which is the optimal throughput in Jellyfish. In all cases, we use 8-edge-disjoint paths between source-destination IP addresses and 8 subflows per MPTCP connection. The traffic pattern is the UT. In DH-Jellyfish, subflows are split over the 4 pairs of IP addresses as explained earlier. We observe that the MPTCP-aware and the random-based approaches perform similarly in Jellyfish, as expected.  In DH-Jellyfish, both approaches perform better. However, the best of the two is the MPTCP-aware approach, which improves the average throughput by 65\% compared to Jellyfish.

\section{Discussion}
\label{sec:discussion}
We have demonstrated that a deterministic subflow routing mechanism requires 33\% and 50\% less subflows than the random approach to achieve equivalent performance for the FatTree and Jellyfish topologies, respectively (see Tables \ref{ft_table}, \ref{jelly_table}). This reduction gives a significant boost to MPTCP's deployment potential in large scale SDN's because it reduces the following: i) the number of installed rules at OpenFlow switches, and ii) the overall load at the SDN controller. Below, we describe these two factors in more detail.

\textbf{(i) Rules on OpenFlow switches:} In large scale SDNs a large number of MPTCP subflows raises scalability concerns because OpenFlow switches can approximately hold mid five figures of rules in their TCAM (64K for commercial switches in \cite{nec_openflow}). Considering a FatTree-based network with $N$ hosts there are $O(N^2)$ possible source-destination pairs. The 33\% reduction of the number of  subflows (4 instead of 6 subflows per MPTCP connection) translates to an order of $2N^2$ fewer subflows in the network. For example, if $N=100$ and there are $4$ instead of $6$ MPTCP subflows between each source-destination pair, then there is a total of $100\times99 \times 2=19800$ fewer subflows in the network (39600 instead of 59400). Moreover, considering a Jellyfish-based network with $N$ hosts the reduction is 50\% (3 instead of 6 subflows per MPTCP connection), which translates to an order of $3N^2$ fewer subflows in the network. Again, if $N=100$ and there are $3$ instead of $6$ MPTCP subflows between each source-destination pair, then there is a total of $100\times99 \times 3=29700$ fewer subflows in the network (29700 instead of 59400). As a result, the required number of rules that need to be installed at the OpenFlow switches reduces significantly.

\textbf{(ii) Load on the SDN controller:} In a reactive SDN environment the controller needs to take action for every subflow. Therefore, reducing the number of subflows without compromising performance is of paramount importance. In our system, we keep things as simple as possible at the controller, which processes subflow setup packets \emph{only}. Specifically, we process these packets in a fast manner using a two-level cache architecture as described in Section \ref{controller_subsection}. In our evaluation, most of the subflow setup packets are processed within 1ms, which is an acceptable time interval. By having an order of $2N^2$-$3N^2$ fewer subflows in the network as explained above, we also significantly reduce the load at the controller. We note that the load at an SDN controller is an inherent issue in the SDN paradigm, and it is suggested that for very large scale SDNs multiple controllers should co-exist to alleviate performance bottlenecks~\cite{hyperflow}.

\section{Conclusion and Future Work} 
\label{conclusion_section}

We have explored an MPTCP-aware SDN approach for deterministically routing MPTCP subflows and better exploiting the available path diversity. Our results show that this approach can provide significant performance benefits in datacenter topologies compared to random-based approaches akin to flow-based ECMP. Our SDN controller accounts for hosts that can have multiple interfaces. As we have shown, sending more than one MPTCP subflow from each interface can improve performance. To facilitate this capability at the hosts, we had to modify the MPTCP implementation. This modification has been included in v0.90 of the MPTCP Linux Kernel implementation~\cite{mptcp_kernel_implementation}. Furthermore, we make our SDN controller publicly available~\cite{mptcp_aware_controller_link}.

All of our evaluation has been performed using long-lived MPTCP flows. As part of our future work, we plan to experiment with scenarios that include a mixture of both short and long-lived flows and study the effect of the proposed approach on flow completion time. Furthermore, we plan to investigate ways for providing deterministic MPTCP subflow routing using a proactive SDN environment.
 

\bibliographystyle{IEEEtran}
\bibliography{IEEEabrv,sig-alternate.bib}

\begin{thebibliography}{10}
\providecommand{\url}[1]{#1}
\csname url@samestyle\endcsname
\providecommand{\newblock}{\relax}
\providecommand{\bibinfo}[2]{#2}
\providecommand{\BIBentrySTDinterwordspacing}{\spaceskip=0pt\relax}
\providecommand{\BIBentryALTinterwordstretchfactor}{4}
\providecommand{\BIBentryALTinterwordspacing}{\spaceskip=\fontdimen2\font plus
\BIBentryALTinterwordstretchfactor\fontdimen3\font minus
  \fontdimen4\font\relax}
\providecommand{\BIBforeignlanguage}[2]{{%
\expandafter\ifx\csname l@#1\endcsname\relax
\typeout{** WARNING: IEEEtran.bst: No hyphenation pattern has been}%
\typeout{** loaded for the language `#1'. Using the pattern for}%
\typeout{** the default language instead.}%
\else
\language=\csname l@#1\endcsname
\fi
#2}}
\providecommand{\BIBdecl}{\relax}
\BIBdecl

\bibitem{al2008scalable}
M.~Al-Fares, A.~Loukissas, and A.~Vahdat, ``{A Scalable, Commodity Data Center
  Network Architecture},'' in \emph{ACM SIGCOMM}, 2008.

\bibitem{guo2009bcube}
C.~Guo, G.~Lu, D.~Li, H.~Wu, X.~Zhang, Y.~Shi, C.~Tian, Y.~Zhang, and S.~Lu,
  ``{BCube: A High Performance, Server-centric Network Architecture for Modular
  Data Centers},'' in \emph{ACM SIGCOMM}, 2009.

\bibitem{greenberg2009vl2}
A.~Greenberg, J.~R. Hamilton, N.~Jain, S.~Kandula, C.~Kim, P.~Lahiri, D.~A.
  Maltz, P.~Patel, and S.~Sengupta, ``{VL2: A Scalable and Flexible Data Center
  Network},'' in \emph{ACM SIGCOMM}, 2009.

\bibitem{singla2012jellyfish}
A.~Singla, C.-Y. Hong, L.~Popa, and P.~B. Godfrey, ``{Jellyfish: Networking
  Data Centers Randomly},'' in \emph{USENIX NSDI}, 2012.

\bibitem{ford2013tcp}
A.~Ford, C.~Raiciu, M.~Handley, and O.~Bonaventure, ``{TCP Extensions for
  Multipath Operation with Multiple Addresses},'' RFC 6824, 2013.

\bibitem{raiciu2011improving}
C.~Raiciu, S.~Barre, C.~Pluntke, A.~Greenhalgh, D.~Wischik, and M.~Handley,
  ``{Improving Datacenter Performance and Robustness with Multipath TCP},'' in
  \emph{ACM SIGCOMM}, 2011.

\bibitem{mptcp_kernel_implementation}
``{MPTCP Linux Kernel Implementation},'' \url{http://www.multipath-tcp.org/}.

\bibitem{hopps2000analysis}
C.~E. Hopps, ``{Analysis of an Equal-Cost Multi-Path Algorithm},'' RFC 2992,
  2000.

\bibitem{al2010hedera}
M.~Al-Fares, S.~Radhakrishnan, B.~Raghavan, N.~Huang, and A.~Vahdat, ``{Hedera:
  Dynamic Flow Scheduling for Data Center Networks.}'' in \emph{USENIX NSDI},
  2010.

\bibitem{mptcp_patch}
``{MPTCP Linux Kernel Patch},''
  \url{https://github.com/multipath-tcp/mptcp/commit/d0f3a6d158d5ce024e4a57f8e6e223e0aa44df2f}.

\bibitem{mptcp_aware_controller_link}
``{MPTCP-aware Controller Implementation},''
  \url{https://github.com/zsavvas/MPTCP-aware-SDN}.

\bibitem{sandri2015benefits}
M.~Sandri, A.~Silva, L.~A. Rocha, and F.~L. Verdi, ``{On the Benefits of Using
  Multipath TCP and Openflow in Shared Bottlenecks},'' in \emph{IEEE AINA},
  2015.

\bibitem{detal2013revisiting}
G.~Detal, C.~Paasch, S.~van~der Linden, P.~Mérindol, G.~Avoine, and
  O.~Bonaventure, ``{Revisiting Flow-Based Load Balancing: Stateless Path
  Selection in Data Center Networks},'' \emph{Computer Networks}, vol.~57,
  no.~5, pp. 1204--1216, April 2013.

\bibitem{agache2015increasing}
A.~Agache, R.~Deaconescu, and C.~Raiciu, ``{Increasing datacenter network
  utilisation with GRIN},'' in \emph{NSDI 15}.

\bibitem{mckeown2008openflow}
N.~McKeown, T.~Anderson, H.~Balakrishnan, G.~Parulkar, L.~Peterson, J.~Rexford,
  S.~Shenker, and J.~Turner, ``{OpenFlow: Enabling Innovation in Campus
  Networks},'' in \emph{ACM SIGCOMM Computer Communication Review}, vol.~38,
  no.~2, 2008, pp. 69--74.

\bibitem{floodlight}
``{Floodlight Controller Documentation },''
  \url{https://floodlight.atlassian.net/wiki/display/floodlightcontroller}.

\bibitem{gill2011understanding}
P.~Gill, N.~Jain, and N.~Nagappan, ``{Understanding Network Failures in Data
  Centers: Measurement, Analysis, and Implications},'' in \emph{ACM SIGCOMM},
  2011.

\bibitem{mininet_paper}
B.~L. et~al., ``{A Network in a Laptop: Rapid Prototyping for Software-defined
  Networks},'' in \emph{ACM HOTNETS}, 2010.

\bibitem{pfaff2009extending}
B.~Pfaff, J.~Pettit, K.~Amidon, M.~Casado, T.~Koponen, and S.~Shenker,
  ``{Extending Networking into the Virtualization Layer},'' in \emph{ACM
  HOTNETS}, 2009.

\bibitem{nec_openflow}
``{NEC PF5248 Datasheet},''
  \url{https://www.necam.com/docs/?id=cd53542b-cc3c-4a2f-adc5-a8fb30ebb74d}.

\bibitem{hyperflow}
A.~Tootoonchian and Y.~Ganjali, ``{HyperFlow: A Distributed Control Plane for
  OpenFlow},'' in \emph{INM/WREN}, 2010.

\end{thebibliography}

\end{document}